\documentclass[aps,amsfonts,nofootinbib]{revtex4-1}
\usepackage{epsfig}
\usepackage{graphicx}
\usepackage{amsmath}
\usepackage{amsbsy}
\usepackage{color}
\usepackage{epsfig}
\usepackage{graphicx}
\usepackage{caption}
\usepackage{subcaption}
\usepackage{amsmath}
\usepackage{amssymb} 
\usepackage{bbm}
\usepackage{amsbsy}
\newcommand{\be}{\begin{equation}}
\newcommand{\ee}{\end{equation}}
\newcommand{\bead}{\begin{aligned}}
\newcommand{\eead}{\end{aligned}}
\newcommand{\1}{\'{\i}}

\newcommand{\bea}{\begin{eqnarray}}
\newcommand{\eea}{\end{eqnarray}}

\usepackage{tikz}
\usepackage{graphicx}
\usepackage{epsfig}
\usepackage{pgfplots}
\usepackage{xparse}
\usepackage{amsbsy}
\usepackage{color}
\usepackage{amsmath}
\usepackage{amssymb}
\usepackage{bbm}
\usepackage{slashed}
\usepackage{caption}
\usetikzlibrary{calc}
\usetikzlibrary{arrows,automata}
\usetikzlibrary{positioning,arrows,patterns,automata}
\usetikzlibrary{decorations}
\usetikzlibrary{decorations.pathmorphing,decorations.markings,trees}

\tikzset{	aphoton/.style={decorate, draw=black},
 	   	photon/.style={decorate, decoration={snake}, draw=black},
		particle/.style={draw=black, postaction={decorate},
        			decoration={markings,mark=at position .5 with {\arrow[draw=black]{>}}}
		},
	    	gluon/.style={decorate, draw=red,
        			decoration={coil,amplitude=4pt, segment length=5pt}},
		vertex/.style={draw,shape=circle,fill=black,minimum size=3pt,inner sep=0pt},
	}

\newcommand{\bb}{\begin{equation}}
\newcommand{\eqb}{\begin{eqnarray}}
\newcommand{\eqf}{\end{eqnarray}}

\def\1{\'{\i}}
\begin{document}
\title{Photons and Dark Photons Through \\ Breit-Wheeler Processes }
\author{Ariel Arza$^{a}$, Jorge Gamboa$^{a}$, and Natalia Tapia$^{a}$} 
\email{ariel.arza@usach.cl,  jorge.gamboa@usach.cl,  natalia.tapiaa@usach.cl}
\affiliation{$^{a}$\ Departamento de  F\'{\i}sica, Universidad de  Santiago de
  Chile, Casilla 307, Santiago, Chile}
\begin{abstract}
A variant of quantum electrodynamics coupled to a dark photon through a kinetic mixing is studied. The analogous of the light-light diagram becomes the conversion process 
$\gamma'\gamma' \to \gamma \gamma$ and an expression for the differential cross section is estimated.  For high energies beams, as in LHC, this differential cross section could be measurable and its magnitude would be typically similar to the total cross section of neutrinos, {\it i.e.} $\sim 10^{-50}\,{\text{m}}^2$.
\end{abstract}
\maketitle
\section{Introduction}
One the most important challenges of particle physics in these days is to explain the origin, formation and dynamics of 
dark matter. Until now even though the existence of dark matter is well established result through the curves of rotation velocity of galaxies and formation of large structures, such as galactic halos and galaxy clusters. However, there are no other sufficiently robust sources corroborating its existence.

According to recent observations of the Fermi Gamma Ray Telescope \cite{fermi} and others \cite{2,3,4,5,6}, an excess of photons is observed which can be explained, on the side, as a consequence of nearby pulsars or remnants of supernovas or, are simply processes of annihilation of dark matter processes in our galaxy.

In the latter context some explanations have been given, for example through the self-interacting dark matter models where the large effective cross section are explained, for example, through Sommerfeld enhancement \cite{spergel, arkani}.

In this note we propose a mechanism that produces an excess of visible photon, which could be attributed to  Breit-Wheeler process \cite{BW} in the sense is showed in the diagram 
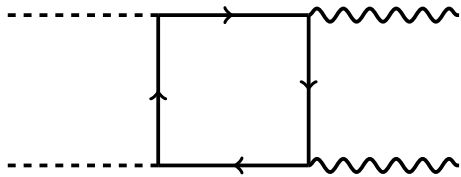
\begin{figure}[h!]
\centering 
\begin{tikzpicture}[node distance=2cm and 2cm]
\coordinate[] (v1);
\coordinate[right=of v1] (v2);
\coordinate[below=of v2] (v3);
\coordinate[left=of v3] (v4);
\coordinate[right=of v2] (v5);
\coordinate[right=of v3] (v6);
\coordinate[left=of v4] (v7);
\coordinate[left=of v1] (v8);
\draw[line width=0.5mm, photon] (v2) -- (v5) node[midway,above=0.1cm] {$$};
\draw[line width=0.5mm, photon] (v3) -- (v6) node[midway,above=0.1cm] {$$};
\draw[line width=0.5mm, dashed, aphoton] (v7) -- (v4) node[midway,above=0.1cm] {$$};
\draw[line width=0.5mm, dashed, aphoton] (v8) -- (v1) node[midway,above=0.1cm] {$$};
\draw[line width=0.5mm, particle] (v1) -- (v2) node[midway,above=0.1cm] {$$};
\draw[line width=0.5mm, particle] (v2) -- (v3) node[midway,above=0.1cm] {$$};
\draw[line width=0.5mm, particle] (v3) -- (v4) node[midway,above=0.1cm] {$$};
\draw[line width=0.5mm, particle] (v4) -- (v1) node[midway,above=0.1cm] {$$};
\end{tikzpicture}
\captionsetup{font=footnotesize}
\captionsetup{justification=raggedright}
\caption{This diagram corresponds to the $\gamma' \gamma' \to \gamma \gamma$ scattering where $\gamma'$ are dark (hidden) photons and $\gamma$ visible ones. }
\label{fig1}
\end{figure} 

Formally this process can be seen as follows; if we cut the diagram as 
\begin{figure}[h!]
\centering 
\begin{tikzpicture}[node distance=1cm and 1cm]
\coordinate[] (v1);
\coordinate[right=of v1] (v2);
\coordinate[right=of v2] (v3);
\coordinate[below=of v3] (v4);
\coordinate[below=of v4] (v5);
\coordinate[left=of v5] (v6);
\coordinate[left=of v6] (v7);
\coordinate[above=of v7] (v8);
\coordinate[left=of v1] (v11);
\coordinate[left=of v11] (v12);
\coordinate[right=of v3] (v31);
\coordinate[right=of v31] (v32);
\coordinate[left=of v7] (v71);
\coordinate[left=of v71] (v72);
\coordinate[right=of v5] (v51);
\coordinate[right=of v51] (v52);
\coordinate[above=of v2] (v21);
\coordinate[below=of v6] (v61);
\draw[line width=0.5mm, photon] (v3) -- (v32) node[midway,above=0.1cm] {$$};
\draw[line width=0.5mm, photon] (v5) -- (v52) node[midway,above=0.1cm] {$$};
\draw[line width=0.5mm, dashed, aphoton] (v12) -- (v1) node[midway,above=0.1cm] {$$};
\draw[line width=0.5mm, dashed, aphoton] (v72) -- (v7) node[midway,above=0.1cm] {$$};
\draw[line width=0.5mm, particle] (v1) -- (v2) node[midway,above=0.1cm] {$$};
\draw[line width=0.5mm, particle] (v2) -- (v3) node[midway,above=0.1cm] {$$};
\draw[line width=0.5mm, particle] (v5) -- (v6) node[midway,above=0.1cm] {$$};
\draw[line width=0.5mm, particle] (v6) -- (v7) node[midway,above=0.1cm] {$$};
\draw[line width=0.5mm, particle] (v3) -- (v5) node[midway,above=0.1cm] {$$};
\draw[line width=0.5mm, particle] (v7) -- (v1) node[midway,above=0.1cm] {$$};
\draw[line width=0.5mm, dashed, red] (v21) -- (v61) node[midway,above=0.1cm] {$$};
\end{tikzpicture}
\captionsetup{font=footnotesize}
\captionsetup{justification=raggedright}
\caption{Note that the {\it cut line} just describes two different physical processes, the left han side is a Breit-Wheeler ones and the right hand side is a pair annihilation. }
\label{fig2}
\end{figure}
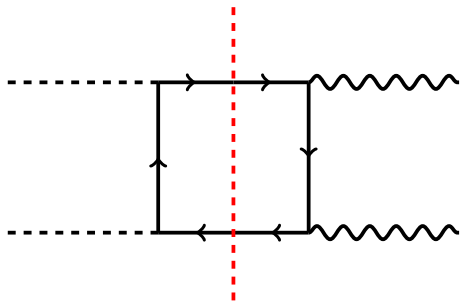 

then --in this mental image-- one can see how the left hand side becomes a Breit-Wheeler process while the right hand side is just an annihilation one, however this {\it constructus} is theoretical and it is useful only to explain how dark photons could be converted to visible photons and viceversa. However in a general context, we need construct a framework where the vertices in FIG. 1 are reproduced and this is another of the goal of this letter. 
\section{The Framework}

Physically, for visible photons, the annihilation and creation of pairs are related by  
\bb 
\sigma_{\text{c}} = 2 |{\bf v}|^2 \, \sigma_{a},
\ee
where $c$ and $a$ denotes creation and annihilations of pairs respectively and ${\bf v}$ is the velocity of the pair. This formula was found by first time by Breit and Wheeler in 1934 although the Breit-Wheeler process $\gamma \gamma \to e^-e^+$ up to now remain as a non-observed prediction of QED. 

The Breit-Wheeler process  is much more difficult to produce than annihilation of pairs one. 
One can note that the equality between cross sections for Breit-Wheeler processes and annihilation pairs could be roughly similar only for extremely energetic massive particles. In \cite{Imp} a photon-photon collider was proposed for detecting Breit-Wheeler process using a high power laser, but so far, there are no conclusive results. 

\medskip 

Basically in a real processes one need to compute the total cross section   
\bb
\sigma_{\text{total}}= \sigma \left(\gamma' \gamma' \to e^+ e^- \to \gamma \gamma\right), \label{total}
\ee
and this formula must show how dark photons are converted in visible ones and the calculation of this observable is another goal of this letter. 
\medskip

In order to implement our ideas let us consider the Lagrangean of QED 

\bb
{\cal  L}_{QED}  =   {\bar  \psi}  \left(  {{   i  \partial \hspace{-.6em}  \slash
      \hspace{.15em}}} +{{ A \hspace{-.5em}  \slash
      \hspace{.15em}}}   -m \right) \psi - \frac{1}{4} F_{\mu \nu} (A) F^{\mu\nu} (A) , 
\label{qed1} 
\ee  
and we add the term
\bb 
{\cal  L}_{int} =  -\frac{1}{4}  F_{\mu  \nu} (B)  F^{\mu  \nu} (B)  +
\frac{\xi}{2} F_{\mu \nu} (B) F^{\mu \nu} (A) + \frac{1}{2}m^2_\gamma B_\mu B^\mu.
\label{dark2}
\ee
where we have added a term corresponding to the kinetic mixing term  $F_{\mu \nu} (B) F^{\mu \nu} (A)$ introduced in \cite{holdom} with $B_\mu$ a dark photon and $m_\gamma$ the mass of this  photon.

\medskip
The total Lagrangean is 
\bb 
{\cal L}_{\text{total}} = {\cal L}_{QED} + {\cal L}_{int}.  \label{total1}
\ee

In order to do tractable this Lagrangean is convenient to diagonalize $A_\mu$ by defining the transformation
\[
A'_\mu = A_\mu - \xi \, B_\mu, \,\,\, \,\,\, B'_\mu = B_\mu,
\] 
(\ref{total1}) becomes 
\bb
{\cal L}_{total} = {\bar  \psi}  \left(  {{   i  \partial \hspace{-.6em}  \slash
      \hspace{.15em}}} +{{ A \hspace{-.4em}  \slash
      \hspace{.15em}}}' +\xi\,{{ B \hspace{-.6em}  \slash
      \hspace{.15em}}} -m \right) \psi - \frac{(1-\xi^2)}{4} F_{\mu \nu} (A') F^{\mu\nu} (A') -\frac{1}{4}  F_{\mu  \nu} (B)  F^{\mu  \nu} (B) +\frac{1}{2}m^2_\gamma B_\mu B^\mu, \label{dark3}
      \ee
  and redefining the field $A'_\mu$ as $\sqrt{1-\xi^2} A'_\mu$  we have 
  \bb
{\cal L}_{total} = {\bar  \psi}  \left(  {{   i  \partial \hspace{-.6em}  \slash
      \hspace{.15em}}} +\frac{1}{\sqrt{1-\xi^2}}{{ A \hspace{-.4em}  \slash
      \hspace{.15em}}}' +\xi\,{{ B \hspace{-.6em}  \slash
      \hspace{.15em}}} -m \right) \psi - \frac{1}{4} F_{\mu \nu} (A') F^{\mu\nu} (A') -\frac{1}{4}  F_{\mu  \nu} (B)  F^{\mu  \nu} (B) +\frac{1}{2}m^2_\gamma B_\mu B^\mu, \label{dark4}
      \ee
  but as $\xi<<1$, we can approximate $\frac{1}{\sqrt{1-\xi^2}} \approx 1$ and therefore (\ref{dark4}) becomes 
   \bb
{\cal L}_{total} = {\bar  \psi}  \left(  {{   i  \partial \hspace{-.6em}  \slash
      \hspace{.15em}}} +{{ A \hspace{-.4em}  \slash
      \hspace{.15em}}} +\xi\,{{ B \hspace{-.6em}  \slash
      \hspace{.15em}}} -m \right) \psi - \frac{1}{4} F_{\mu \nu} (A) F^{\mu\nu} (A) -\frac{1}{4}  F_{\mu  \nu} (B)  F^{\mu  \nu} (B) +\frac{1}{2}m^2_\gamma B_\mu B^\mu, \label{dark4}
      \ee
    and the primes have been removed by simplicity. 
    \medskip
    
    Now we note that at fourth-order in perturbation theory the diagram FIG. \ref{fig1} is reproduced where the vertices are 
 $-i\gamma_\mu$ and $-i\,\xi\, \gamma_\mu$ for visible and dark photons respectively. 
 \medskip
 
 Taking this fact in mind, the differential cross section per unit solid angle $\Omega$, in the limit of low energy, \cite{karplus}  for FIG. \ref{fig1} is    
\bb
\left.\frac{d\sigma(s,\theta)}{d\Omega}\right|_{\gamma'\gamma'\to e^-e^+\to\gamma\gamma}=\kappa\frac{\alpha^4\xi^4}{64\pi^2m^8}s^3(3+\cos^2\theta)^2, 
\ee 
where $\alpha= e^2/4\pi$ is the fine structure constant, $\kappa$ is a coefficient that depend of the explicit (difficult) calculation, but that is not important for this analysis and $s$ is the Mandelstam variable  
\[
s= \left(p_1+p_2\right)^2, 
\] 
where $p_{1,2}$ are initial dark photons momenta.

In the center of mass frame, $s$ can be written as 
\[
s=4\left( {\bf p}^2+m_\gamma^2 \right) \equiv 4\, E^2 ,
\] 
with $E=\sqrt{{\bf p}^2+m_\gamma^2}$ the total energy of the dark photons.
\medskip 

Thus, we find
\bb
\left.\frac{d\sigma(E,\theta)}{d\Omega}\right|_{\gamma'\gamma'\to e^-e^+\to\gamma\gamma}=\kappa\frac{\alpha^4\xi^4}{\pi^2m^2}\left(\frac{E}{m}\right)^6(3+\cos^2\theta)^2. \label{dark11}
\ee

For the high energy limit ($E\gg m$), we do not have an analytical expression for differential cross section such as (\ref{dark11}), however it can be approached for $\theta=0$ and $\theta=\pi/2$. We have
\bb
\left.\frac{d\sigma(E,0)}{d\Omega}\right|_{\gamma'\gamma'\to e^-e^+\to\gamma\gamma}\sim\frac{\alpha^4\xi^4}{\pi^2m^2}\left(\frac{m}{E}\right)^2\left(\ln\frac{E}{m}\right)^4 \label{dark12}
\ee
and
\bb 
\left.\frac{d\sigma(E,\pi/2)}{d\Omega}\right|_{\gamma'\gamma'\to e^-e^+\to\gamma\gamma}\sim\frac{\alpha^4\xi^4}{\pi^2m^2}\left(\frac{m}{E}\right)^2. \label{dark13}
\ee

\begin{figure}[h!]
\centering
\begin{tikzpicture}[node distance=1cm and 1cm]
\coordinate[] (v1);
\coordinate[right=of v1] (v2);
\coordinate[right=of v2] (v3);
\coordinate[below=of v3] (v4);
\coordinate[below=of v4] (v5);
\coordinate[left=of v5] (v6);
\coordinate[left=of v6] (v7);
\coordinate[above=of v7] (v8);
\coordinate[left=of v1] (v11);
\coordinate[left=of v11] (v12);
\coordinate[right=of v3] (v31);
\coordinate[right=of v31] (v32);
\coordinate[left=of v7] (v71);
\coordinate[left=of v71] (v72);
\coordinate[right=of v5] (v51);
\coordinate[right=of v51] (v52);
\coordinate[above=of v2] (v21);
\coordinate[below=of v6] (v61);
\draw[line width=0.5mm, photon] (v3) -- (v32) node[midway,above=0.1cm] {$$};
\draw[line width=0.5mm, photon] (v5) -- (v52) node[midway,above=0.1cm] {$$};
\draw[line width=0.5mm, photon] (v12) -- (v1) node[midway,above=0.1cm] {$$};
\draw[line width=0.5mm, dashed, aphoton] (v72) -- (v7) node[midway,above=0.1cm] {$$};
\draw[line width=0.5mm, particle] (v1) -- (v2) node[midway,above=0.1cm] {$$};
\draw[line width=0.5mm, particle] (v2) -- (v3) node[midway,above=0.1cm] {$$};
\draw[line width=0.5mm, particle] (v5) -- (v6) node[midway,above=0.1cm] {$$};
\draw[line width=0.5mm, particle] (v6) -- (v7) node[midway,above=0.1cm] {$$};
\draw[line width=0.5mm, particle] (v3) -- (v5) node[midway,above=0.1cm] {$$};
\draw[line width=0.5mm, particle] (v7) -- (v1) node[midway,above=0.1cm] {$$};
\draw[line width=0.5mm, dashed, red] (v21) -- (v61) node[midway,above=0.1cm] {$$};
\end{tikzpicture}
\ \ \ \ \ \ \ \
\begin{tikzpicture}
\coordinate[] (v1);
\coordinate[right=of v1] (v2);
\coordinate[right=of v2] (v3);
\coordinate[below=of v3] (v4);
\coordinate[below=of v4] (v5);
\coordinate[left=of v5] (v6);
\coordinate[left=of v6] (v7);
\coordinate[above=of v7] (v8);
\coordinate[left=of v1] (v11);
\coordinate[left=of v11] (v12);
\coordinate[right=of v3] (v31);
\coordinate[right=of v31] (v32);
\coordinate[left=of v7] (v71);
\coordinate[left=of v71] (v72);
\coordinate[right=of v5] (v51);
\coordinate[right=of v51] (v52);
\coordinate[above=of v2] (v21);
\coordinate[below=of v6] (v61);
\draw[line width=0.5mm, photon] (v3) -- (v32) node[midway,above=0.1cm] {$$};
\draw[line width=0.5mm, photon] (v5) -- (v52) node[midway,above=0.1cm] {$$};
\draw[line width=0.5mm, dashed, aphoton] (v12) -- (v1) node[midway,above=0.1cm] {$$};
\draw[line width=0.5mm, photon] (v72) -- (v7) node[midway,above=0.1cm] {$$};
\draw[line width=0.5mm, particle] (v1) -- (v2) node[midway,above=0.1cm] {$$};
\draw[line width=0.5mm, particle] (v2) -- (v3) node[midway,above=0.1cm] {$$};
\draw[line width=0.5mm, particle] (v5) -- (v6) node[midway,above=0.1cm] {$$};
\draw[line width=0.5mm, particle] (v6) -- (v7) node[midway,above=0.1cm] {$$};
\draw[line width=0.5mm, particle] (v3) -- (v5) node[midway,above=0.1cm] {$$};
\draw[line width=0.5mm, particle] (v7) -- (v1) node[midway,above=0.1cm] {$$};
\draw[line width=0.5mm, dashed, red] (v21) -- (v61) node[midway,above=0.1cm] {$$};
\end{tikzpicture}
\captionsetup{font=footnotesize}
\captionsetup{justification=raggedright}
\caption{This diagram corresponds to the $\gamma'\gamma' \to \gamma \gamma$ scattering where $\gamma'$ are dark (hidden) photons and $\gamma$ visible ones. }
\label{fig3}
\end{figure}
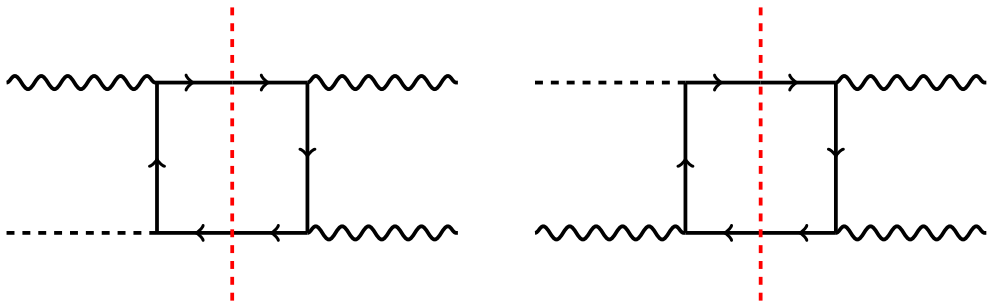

Both cases (high and low energy limits) are very interesting because show how, in principle, dark photons could be transformed into visibles ones. It is also important to mention the process $\gamma'\gamma\to\gamma\gamma$ (see FIG. \ref{fig3}), where the initial photon can be prepared in the laboratory. In this case the formulas of differential cross sections should be scale as $d\sigma/d\Omega\sim\xi^2$. For instance, the high energy limit at $\theta=0$ is given by
\bb
\left.\frac{d\sigma(E,0)}{d\Omega}\right|_{\gamma'\gamma'\to e^-e^+\to\gamma\gamma}\sim\frac{\alpha^4\xi^2}{\pi^2m^2}\left(\frac{m}{E}\right)^2\left(\ln\frac{E}{m}\right)^4 \label{dark14}
\ee

\begin{figure}[h!]
    \includegraphics[scale=1.0]{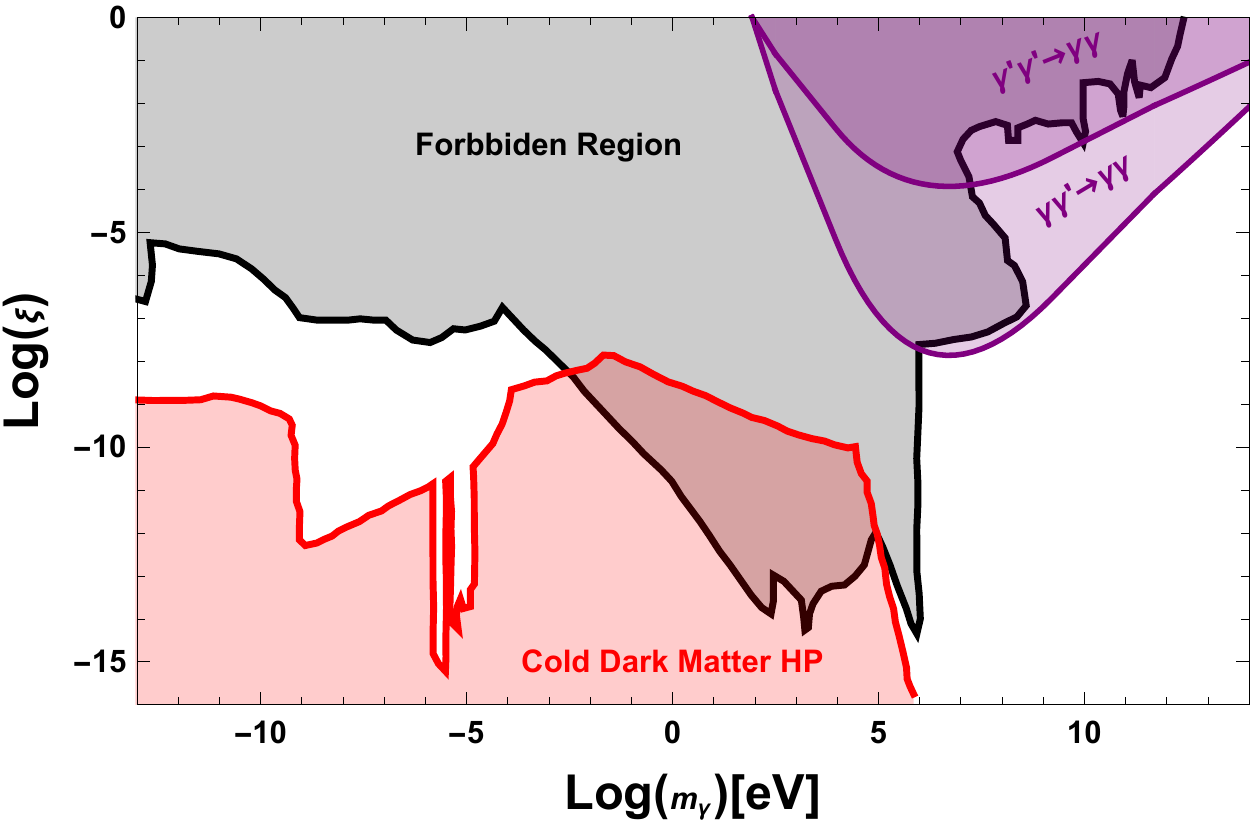}
    \captionsetup{justification=raggedright}
    \captionsetup{font=footnotesize}
  \caption{This is an exclusion plot in search for hidden photons, the purple region is where the process studied  is allowed for the processes $\gamma'\gamma'\to\gamma\gamma$ and $\gamma'\gamma\to\gamma\gamma$ with a differential cross section of $10^{-50}\text{m}^2$ at $\theta=0$. Gray region correspond to the excluded space and the pink region is where Hidden Photons can be cold dark matter.}
  \label{grafico}
\end{figure}

 FIG. \ref{grafico} shows, for the case $m_\gamma\gg|{\bf p}|$, the space of parameters $(m_\gamma,\xi)$ where this process is allowed, taking into account typically like-neutrinos measurable cross sections \cite{neutrinos}. 
\medskip 

Notice that below the black line and for masses $m_\gamma >10^6$\, eV and $\xi >10^{-7}$, there is a window  where differential cross sections of the order of $d\sigma/d\Omega\sim10^{-50}\, \text{m}^2$ are allowed, which are typically values for neutrino-like particles. We found that this effect is sensitive in this unexplored space of parameters, where the masses $m_\gamma$ are very large. Beyond these values, corrections of the standard model must be included. 

\section{Comments and Conclusion}
There are several strategies aimed at detecting dark photons, for example using the bremsstrahlung where processes such as $e^-Z \to e^- Z A'$ where $Z$ is nuclear target and $A'$ is produced very forward and carrying the most of the beam energy, another example is the $e^-e^+ \to \gamma A'$ where $A'$ is a difference of energy which is attributed to dark photons and so on (these examples are explained in \cite{rev}). 

In this paper we have considered a different route and we have proposed a mechanism where the basic process is  a Breit-Wheeler one with two initial dark photons or one dark photon with a ordinary photon  --including virtual processes-- that convert this photons in visibles ones. 

Although Breit-Wheeler processes are still not observed even in conventional quantum electrodynamics, we can speculate with the orders of magnitude of possible cross sections. 
Indeed, the total cross section (\ref{dark12}) can be written as 
\eqb
\left.\frac{d\sigma(E,0)}{d\Omega}\right|_{\gamma'\gamma'(\gamma)\to e^-e^+\to\gamma\gamma}\ &\sim \xi^{4(2)}\times \left(\frac{E}{m}\right)^6\times 10^{-33} \text{m}^2,&\ \ \ \ \ \ \ \ \ E\ll m, \nonumber
\\
\left.\frac{d\sigma(E,0)}{d\Omega}\right|_{\gamma'\gamma'(\gamma)\to e^-e^+\to\gamma\gamma}&\sim\xi^{4(2)}\left(\frac{m}{E}\right)^2\left(\ln\frac{E}{m}\right)^4\times 10^{-34}\text{m}^2,&\ \ \ \ \ \ \ \ \ E\gg m. \label{dark15}
\eqf

In both cases in (\ref{dark15}) the low and high energy limit are considered but the last one is more important here because there is a  parameters space still  unexplored.  For example in LHC can be produced --probably as secondary photons-- with 1 TeV of energy, then if $\omega \sim 1\text{ TeV}$ could be possible cross sections comparable to the neutrino scattering ones \cite{neutrinos}, however this last fact needs a more detailed analysis.  
\medskip  

This work was supported by FONDECYT/Chile grants 1130020 (J.G.),  N.T. thanks to the  
Conicyt fellowship 21160064 and USA-1555 (A.A.).  We would like to  thank Prof. F. M\'endez for useful discussions and  Prof. S. L. Adler by suggesting us the reference \cite{karplus}.

\end{document}